\documentclass[aps,prl,twocolumn,showpacs,preprintnumbers,amsmath,amssymb,superscriptaddress,floatfix]{revtex4}
\usepackage{amsmath,amssymb,amsfonts,amsthm}
\usepackage[ascii]{inputenc}
\usepackage{bm}
\usepackage{graphicx}
\usepackage{wasysym}
\usepackage{framed}
\usepackage{tikz}

\usepackage[normalem]{ulem}
\usetikzlibrary{trees}
\graphicspath{{figures/}}

\begin{document}

\title{Paradeisos: a perfect hashing algorithm for many-body eigenvalue problems}

\author{C.~J.~Jia}
\altaffiliation{C.~J.~Jia and Y.~Wang contributed equally to this work. Correspondence should be addressed to C.~J.~Jia (chunjing@stanford.edu) or Y.~Wang (yaowang@stanford.edu).}
\affiliation{Stanford Institute for Materials and Energy Sciences, SLAC National Accelerator Laboratory, 2575 Sand Hill Road, Menlo Park, CA 94025}

\author{Y.~Wang}
\altaffiliation{C.~J.~Jia and Y.~Wang contributed equally to this work. Correspondence should be addressed to C.~J.~Jia (chunjing@stanford.edu) or Y.~Wang (yaowang@stanford.edu).}
\affiliation{Stanford Institute for Materials and Energy Sciences, SLAC National Accelerator Laboratory, 2575 Sand Hill Road, Menlo Park, CA 94025}
\affiliation{Departments of Applied Physics, Stanford University, CA 94305}

\author{C.~B.~Mendl}
\affiliation{Stanford Institute for Materials and Energy Sciences, SLAC National Accelerator Laboratory, 2575 Sand Hill Road, Menlo Park, CA 94025}

\author{B.~Moritz}
\affiliation{Stanford Institute for Materials and Energy Sciences, SLAC National Accelerator Laboratory, 2575 Sand Hill Road, Menlo Park, CA 94025}
\affiliation{Department of Physics and Astrophysics, University of North Dakota, Grand Forks, ND 58202, USA}

\author{T.~P.~Devereaux}
\affiliation{Stanford Institute for Materials and Energy Sciences, SLAC National Accelerator Laboratory, 2575 Sand Hill Road, Menlo Park, CA 94025}
\affiliation{Geballe Laboratory for Advanced Materials, Stanford University, CA 94305}

\date{\today}

\begin{abstract}
We describe an essentially perfect hashing algorithm for calculating the position of an element in an ordered list, appropriate for the construction and manipulation of many-body Hamiltonian, sparse matrices.  Each element of the list corresponds to an integer value whose binary representation reflects the occupation of single-particle basis states for each element in the many-body Hilbert space.  The algorithm replaces conventional methods, such as binary search, for locating the elements of the ordered list, eliminating the need to store the integer representation for each element, without increasing the computational complexity.  Combined with the ``checkerboard'' decomposition of the Hamiltonian matrix for distribution over parallel computing environments, this leads to a substantial savings in aggregate memory.  While the algorithm can be applied broadly to many-body, correlated problems, we demonstrate its utility in reducing total memory consumption for a series of fermionic single-band Hubbard model calculations on small clusters with progressively larger Hilbert space dimension.
\end{abstract}

\maketitle

\section{Introduction}
A number of wavefunction- and Green's function-based numerical techniques have been developed to address the problem of strongly interacting electrons in lattice models for condensed matter systems.~\cite{FehskeBook2008}  Among the most widely and straightforwardly applied methods is small cluster exact diagonalization, which is particularly well suited to problems with strong interactions where the important physics remains local or relatively short-ranged.  One explicitly constructs a many-body Hamiltonian from the full Hilbert space that consists of all allowed multi-particle configurations of single-particle states for the full lattice problem.~\cite{DagottoRMP}  The usefulness of this technique is limited by the exponential growth of the Hilbert space dimension $D$ with increasing size of the clusters.

Full diagonalization of the Hamiltonian -- evaluating all eigenvalues and eigenvectors -- remains impractical for all but the smallest problems with a typical computational complexity of $O(D^{3})$.  However, accurate information about the ground (lowest energy) state and several low lying excited states can be sufficient for zero, or low temperature properties of the model.  To that end, iterative Krylov subspace methods, such as Lanczos or Arnoldi~\cite{golub2012matrix}, can be employed to reduce the computational burden. 
Dynamical properties can be evaluated from the ground (and excited) state eigenfunctions using secondary numerical methods~\cite{jia2015using, wang2014real}, such as the continued fraction expansion~\cite{CFE} or bi-conjugate gradient stabilized techniques~\cite{BiCGS}. 

A reduction of the computational complexity for these methods also comes from the fact that the Hamiltonian matrix is typically sparse, with a polynomial number of non-zero elements (usually only several hundred non-zero elements) per row or column, typically orders of magnitude smaller than the Hilbert space dimension $D$.  
However, exponential growth of the Hilbert space means that the Hamiltonian matrix or even the wavefunctions might be too large to store in the memory available on a single compute node.  This is certainly the case for the largest and most challenging problems for exact diagonalization.  The nature of modern parallel computing environments with a large number of lightweight cores per node and limited memory per core (or per node) makes the issue of efficient memory utilization in exact diagonalization one of the most significant bottlenecks to performance and scalability.     

\begin{figure}[tb]
\centering
\includegraphics[width=0.8\columnwidth]{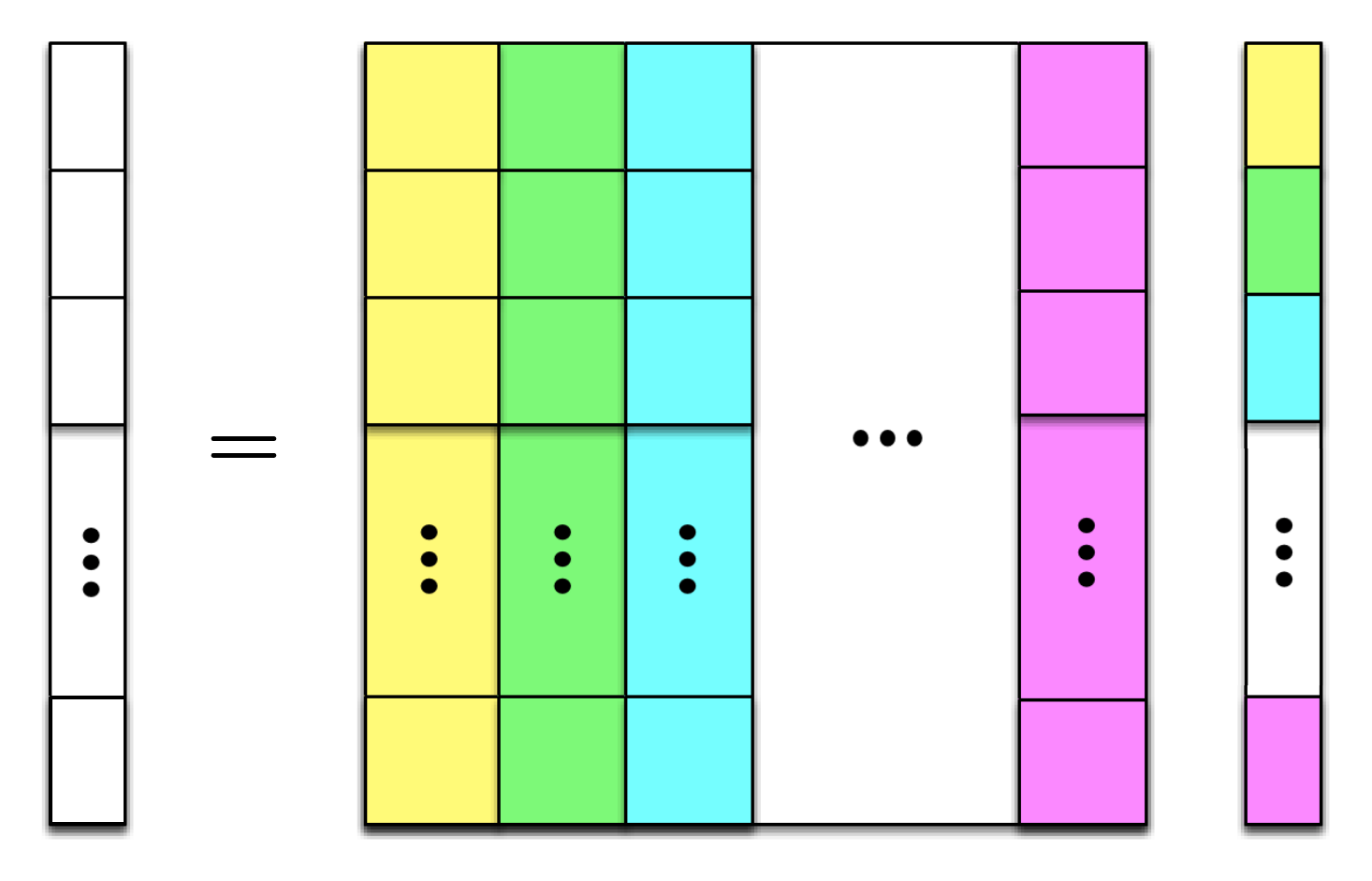}
\caption{Checkerboard decomposition of the Hamiltonian matrix. The colors represent a range of indices in the many-body basis, for both the sparse matrix and wavefunctions, distributed between different compute nodes.}
\label{fig:checkerboard}
\end{figure}
One solution to this problem has been a ``checkerboard'' decomposition of the Hamiltonian matrix as illustrated in Fig.~\ref{fig:checkerboard}.  This decomposition distributes a different partition of the matrix and wavefunctions (blocks with the same color) to different processors, or compute nodes, to balance the memory distribution. When combined with graph partitioning and matrix reordering algorithms, such as those in the PARMETIS library~\cite{METIS}, this scheme maximizes data locality, reducing parallel communications.  
The remaining performance and 
memory bottleneck comes from the construction of the matrix from the many-body basis (Hilbert space) and the one-to-one mapping between each state and the appropriate index.  A traditional binary search algorithm may be employed when the states are properly ordered; however, this method usually requires a hash table, which encodes the index-state map, resident in memory for each processor, or compute node, to minimize communications overhead.  An ideal hashing scheme would allow for the evaluation of the index-state map, in both the forward and reverse directions, without any memory overhead.  

In this paper, we effectively devise such a perfect hashing algorithm, \emph{Paradeisos}, to replace conventional search with a direct mapping. This eliminates the need to store a table for the mapping between the vector-matrix index and the many-body basis.  The method becomes particularly advantageous for large problem sizes and in parallel computing environments where the ``checkerboard'' decomposition must be used for distributed matrix-vector storage. 
We describe \emph{Paradeisos} in the framework of a single-band, or single-orbital, model for two species of interacting fermions (``spin-up'' $\uparrow$ and ``spin-down'' $\downarrow$) on a lattice.  Extension of this algorithm to multi-orbital models, or models for restricted bosonic systems, can be accomplished with straightforward modifications. 
The paper is organized as follows: first, we introduce the algorithm and the forward and backward mapping in the general case; second, we discuss the implementation of the algorithm in symmetry projected subspaces of the full Hilbert space; and finally, we explore the performance of the algorithm and its scaling with the number of parallel processes. 

\section{Hilbert space and model Hamiltonian}
\label{sec:basis_construction}

\subsection{Many-body basis}

Start with the single-site fermionic problem. 
There are four configurations due to the Pauli exclusion principle: $\lvert 0 \rangle$, $\lvert\uparrow\rangle$, $\lvert\downarrow\rangle$, $\lvert\uparrow \downarrow \rangle$ (or in occupation and spin direct product basis $\lvert 0 \rangle_\uparrow \otimes \lvert 0 \rangle_\downarrow $, $\lvert 1 \rangle_\uparrow \otimes \lvert 0 \rangle_\downarrow $, $\lvert 0 \rangle_\uparrow \otimes \lvert 1 \rangle_\downarrow $, $\lvert 1 \rangle_\uparrow \otimes \lvert 1 \rangle_\downarrow$). For a multi-site problem, 
one typically works in the canonical ensemble with fixed total electron number. In general, for an $N$-site cluster with $n_\uparrow$ spin-up and $n_\downarrow$ spin-down electrons, the Hilbert space dimension
$D = C_N^{n_\uparrow} \cdot C_N^{n_\downarrow}$, where $C_m^n$ is the binomial coefficient.  

Any basis element of the Hilbert space, represented in terms of the occupation of single-particle states, can be constructed from the vacuum $\vert 0 \rangle$ by repeated application of creation operators $c^\dagger_{i\sigma}$, where $\sigma \in \{\uparrow, \downarrow\}$ and $i \in \{0, 1, \ldots, N-1 \}$ indexes lattice positions, such that
\[\lvert \varphi \rangle = \underbrace{\ldots c^\dagger_{i\uparrow} \ldots}_{n_\uparrow} \underbrace{\ldots c^\dagger_{j\downarrow} \ldots}_{n_\downarrow} \lvert 0 \rangle. \]
Note the ``normal ordering'' convention with spin-down operators on the right, and we take the increasing site index right-to-left.  The convention ensures proper
antisymmetrization for fermion exchange.  As an illustration to which we will return throughout the discussion of the algorithm, consider a simple 4-site lattice model with two spin-up ($n_\uparrow = 2$) and three spin-down ($n_\downarrow = 3$) electrons.  In this case, $D=24$ and one particular element of the Hilbert space basis
\[c^\dagger_{2\uparrow} c^\dagger_{0\uparrow} c^\dagger_{3\downarrow} c^\dagger_{1\downarrow} c^\dagger_{0\downarrow} \lvert 0\rangle = (0 \uparrow 0 \uparrow) \otimes (\downarrow 0 \downarrow \downarrow).\]

For practical computations, all basis elements can be enumerated by noting the correspondence of the occupation and spin direct product representation with bit sequences for unique integer values. 
Turn again to the example of one particular basis element for the 4-site lattice model.
\[(0 \uparrow 0 \uparrow) \otimes (\downarrow 0 \downarrow \downarrow) \to 0101_2 \otimes 1011_2,\]
where each occurrence of $\uparrow$ or $\downarrow$ has been replace by a $1$ (occupied) with $0$ otherwise.  The subscript $2$ on each sequence denotes binary notation for clarity.
A lexicographical binary representation can be constructed separately for every spin-up and spin-down element comprising the Hilbert space basis.  For the 4-site example, results are presented in the following table.
\begin{center}
\begin{tabular}{c | l || c | l}
index & $\uparrow$ element & index & $\downarrow$ element \\
\hline
0 & $0011_2 =  3$ & 0 & $0111_2 =  7$ \\
1 & $0101_2 =  5$ & 1 & $1011_2 = 11$ \\
2 & $0110_2 =  6$ & 2 & $1101_2 = 13$ \\
3 & $1001_2 =  9$ & 3 & $1110_2 = 14$ \\
4 & $1010_2 = 10$ &   & \\
5 & $1100_2 = 12$ &   & \\
\end{tabular}
\end{center}
In general, the lexicographical next state, 
advancing the index $p \rightarrow p+1$,
can be determined using just a few bit operations ~\cite{FermiFab2011, AndersonWebpage}. A tensor product of the spin-up and spin-down contributions defines the complete basis.  This tensor product structure typically breaksdown within restricted symmetry subspaces, which we will address in a subsequent section. 

\subsection{Many-body Hamiltonian}

Although the algorithm does not depend on the details of the many-body Hamiltonian, for simplicity, we consider the single-band Hubbard model. Note that the symmetry of the Hamiltonian affects the selection of the subspace.
The single-band Hubbard Hamiltonian $H$ includes both kinetic and on-site Coulomb interaction terms, written conventionally as
\begin{eqnarray}
\label{eq:hubbard}
H &=& H_K + H_I \nonumber\\
  &=& -\sum_{ij, \sigma}\big( t_{ij} c^\dagger_{i\sigma} c_{j\sigma} + h.c. \big) + U \sum_{i} n_{i\uparrow} n_{i\downarrow},
\end{eqnarray}
where $n_{i\sigma} = c^\dagger_{i\sigma} c_{i\sigma}$ is the standard number operator. The kinetic terms include ``hopping'' of electrons on the lattice with an energy $t_{ij}$, typically restricted to nearest or next-nearest neighboring sites, and $U$ parameterizes the on-site interaction strength. 

$H$ can be viewed as a one-to-many map of the Hilbert space into itself.  Consider first the interaction terms.  In a representation based on real-space occupation on the lattice, the interaction depends on the ``double occupancy'' -- how many sites have occupation by both $\uparrow$ and $\downarrow$ fermions.  Comprised solely of number operators, this term represents a map of each many-body Hilbert space basis element to itself with a prefactor.  In bit operations, the ``double occupancy'' (and by extension the Hamiltonian matrix elements) can be computed from the bitwise {\bf \texttt{AND}} between the $\uparrow$ and $\downarrow$ parts of the direct product. 

The kinetic terms represent a more complicated map.  The movement of fermions on the lattice, or ``hopping'', mixes basis elements through a change in the fermion occupation.  One must determine not only the weights for the mapping (Hamiltonian matrix elements), but also identify the index (or indices) within the basis to which $H$ maps each element.  Return to our example of the 4-site lattice; and consider a kinetic term with only nearest-neighbor hopping $t$ and sites linked together cyclically, as in a one-dimensional loop ($0 \leftrightarrow 1 \leftrightarrow 2 \leftrightarrow 3 \leftrightarrow 0$).  For simplicity, we address only the spin-up kinetic term acting on our test element $0101_2 \otimes 1011_2$, where the first bit sequence of the product corresponds to spin-up.  In this case,
\begin{align}
 H_K^{\uparrow} & (0101_2 \otimes 1011_2) = \nonumber\\
-t &\big( 1001_2 + 0011_2 + 0110_2 - 1100_2 \big) \otimes 1011_2,\nonumber
\end{align}
where the minus sign in front of $1100_2$ results from the normal ordering convention and the antisymmetry of fermion exchange.  One can immediately read-off the index for each state from the table created for this example; however, ideally one would like a more automated and less brute force method for determining the new indices under this mapping for a state-of-art size of quantum many-body problem.

\subsection{Index determination}

One may employ conventional binary search algorithm~\cite{BinarySearch}, to determine the indices to which a function such as $H$ will map each basis element. The binary search algorithm has time complexity $\mathcal{O}(\ln D)$. Since $D=C_N^{n_\uparrow} \cdot C_N^{n_\downarrow} < 4^N$, an upper bound for the time complexity of binary search is $\mathcal{O}(\ln 4^N)$ or $\mathcal{O}(N)$. 
However, one must typically store the enumerated bit sequences for all basis elements in memory to reduce the communications overhead.

In the following sections we describe \emph{Paradeisos}, a direct ``forward mapping'' from a bit sequence to its basis index, mitigating the need for search, and by extension storage.  Crucially, this algorithm has the same time complexity as traditionally binary search methods. 
When combined with advanced matrix decomposition techniques for parallel computing environments, \emph{Paradeisos} has the potential to significantly reduce aggregate memory consumption and improve performance (see the section on ``Numerical Benchmarks").  

\section{Mapping functions}
\label{sec:mapping_functions}
The key feature of \emph{Paradeisos} is a mapping from a bit sequence to an index. As opposed to binary search, which identifies an index from an ordered list through an iterative series of bisections, \emph{Paradeisos} determines the (absolute) index of an arbitrary bit sequence of the list by mathematically evaluating the total number of different configurations between it and a known pattern. 

First, we define the distance between two elements $\varphi_1$ and $\varphi_2$ by the difference of their indices $\mathrm{dist}\{\varphi_1,\varphi_2\} = \mathrm{idx}(\varphi_2)-\mathrm{idx}(\varphi_1)$.
However, directly evaluating this distance is non-trivial. It can be straightforward in the special case of \emph{simple neighbors}
, where two bit sequences differ by exchange of a single 1-bit across consecutive 0-bits. As an example consider two elements, each with 4 non-zero \emph{least significant bits} (LSBs) up to the position of this exchange.  The first 
\begin{equation*}
\begin{split}
|\varphi_1\rangle := \,&\ldots \overset{j_2}{0}\,0\ 0\underbrace{\overset{j_1}{{\bf {\bf 1}}}\,0\ 0\ 1\ 0\ 1\, \overset{i=0}{1_2}}_{n = 4} \\
\end{split}
\end{equation*}
has an index $\mathrm{idx}(\varphi_1)=p$, and the second
\begin{equation*}
\begin{split}
|\varphi_2\rangle :=\,&\dots \underbrace{\overset{j_2}{{\bf 1}}\,0\ 0\,\overset{j_1}{0}\ 0\ 0\ 1\ 0\ 1\, \overset{i=0}{1_2}}_{n = 4}
\end{split}
\end{equation*}
has an index $\mathrm{idx}(\varphi_2)=q$.  The distance between these two bit sequences depends on the number of configurations of the $n$ bits between them, which is given simply by 
\begin{equation}
\label{eq:index_free_hopping}
\mathrm{dist}\{\varphi_1,\varphi_2\} = q - p = C_{j_2}^n - C_{j_1}^n,
\end{equation}
with a convention that the bit position starts from $i=0$ and that $C_m^n = 0$ for $m < n$. 
Note that Eq.~\eqref{eq:index_free_hopping} is independent of the particular bit configuration to the right of $j_1$, nor does it depend upon the configuration to the left of $j_2$ if the 
total number of 1-bits were greater than $n$.
One can construct the mapping between a bit sequence and its corresponding index, via successive exchanges from a given sequence with a known index, by taking advantage of the distance between simple neighbors.  By construction, index $0$ for the first element in a list of $N$-bit numbers with $n$ non-zero bits ($n < N$) belongs to the bit sequence with the $n$ LSBs set to $1$:
\begin{equation*}
\begin{split}
\lvert \varphi_0\rangle :=& \underbrace{\overset{N-1}{0} \ldots\ \overset{n}{0}\ \underbrace{\overset{n-1}{1}\ \ldots \overset{i = 0}{1_2}}_{n}}_{N}, \\
\mathrm{idx}(\varphi_0) &= 0 
\end{split}
\end{equation*}
For any other $N$-bit sequence $\lvert \varphi \rangle$ in the ordered list, its $n$ 1-bits will occupy positions $j_{n-1} > j_{n-2} > \ldots > j_{1} > j_{0}$.  Starting from $\lvert \varphi_0 \rangle$, one can construct a sequence of $n$ simple neighbors $\{\lvert \varphi_0^{(n)} \rangle, \dots, \lvert \varphi_0^{(1)} \rangle = \lvert \varphi \rangle\}$ by sequential exchange of the $m^{\mathrm{th}}$ LSB, $\{m=n, \ldots, 1\}$.  Thus,
\begin{equation*}
\begin{split}
\lvert \varphi_0\rangle :=&\ \overset{N-1}{0} \ldots\ \overset{n}{0}\ \overset{n-1}{1}\ \ldots \overset{i = 0}{1_2} \\
\lvert \varphi_0^{(n)}\rangle :=&\ \ldots \overset{j_{n-1}}{1} \ldots\ \overset{n-1}{0}\ \overset{n-2}{1} \ldots \overset{i = 0}{1_2} \\
\lvert \varphi_0^{(n-1)}\rangle :=&\ \ldots \overset{j_{n-1}}{1} \ldots\ \overset{j_{n-2}}{1} \ldots \overset{n-2}{0}\ \overset{n-3}{1} \ldots \overset{i = 0}{1_2} \\
\vdots& \\
\lvert \varphi_0^{(2)}\rangle :=&\ \ldots \overset{j_{n-1}}{1} \ldots\ \overset{j_{n-2}}{1} \ldots \overset{j_{1}}{1} \ldots \overset{1}{0}\ \overset{i = 0}{1_2} \\
\lvert \varphi \rangle = \lvert \varphi_0^{(1)}\rangle :=&\ \ldots \overset{j_{n-1}}{1} \ldots\ \overset{j_{n-2}}{1} \ldots \overset{j_{1}}{1} \ldots \overset{j_{0}}{1} \dots_2,
\end{split}
\end{equation*}
and one can now read-off the index simply as the accumulated distance between simple neighbors
\begin{equation}\label{eq:idx}
\mathrm{idx}(\varphi) = C_{j_0}^1 + C_{j_1}^2 + \dots + C_{j_{n-2}}^{n-1} + C_{j_{n-1}}^n.
\end{equation}

\noindent
\begin{tabular}{|l|}
\hline
\textbf{\emph{Paradeisos} forward map}\\
Goal: For $n$ electrons and $N$ sites, 
map $\varphi \mapsto \textrm{idx}(\varphi)$\\
\hline
\textbf{function} {\it forward\_map}\,($\varphi$)\\
\indent \textbf{declare} $\mathrm{idx}$ = 0, $m$ = 0.\\
\indent \textbf{for} $i = 0$ to $N-1$:\\
\indent \indent \textbf{if} $i$-th bit in $\varphi$ is \texttt{TRUE}:\\
\indent \indent \indent $m$\ $\leftarrow$\ $m + 1$\\
\indent \indent \indent $\mathrm{idx}$\ $\leftarrow$\ $\mathrm{idx} + C_{i}^m$\\
\indent \indent \textbf{end}\\
\indent \textbf{end}\\
\indent \textbf{return} $\mathrm{idx}$ \\
\hline
\end{tabular}
\newline

\noindent For efficiency, the binomial coefficients $C_i^m$ for $i, m \in \{0, 1, \dots, N-2, N-1\}$ can be precomputed recursively $C_i^m = C_{i-1}^m + C_{i-1}^{m-1}$, and stored in a lookup-table. The time complexity for {\it Paradeisos} is O($N$), comparable to traditional binary search. 

As an example, return to the problem of 2 spin-up and 3 spin-down electrons on a 4 site lattice and, for simplicity, consider only element $1100_2$ in the spin-up part of the Hilbert space basis.
After exchange of the highlighted (boldface/underline) bits,
\begin{equation*}
\begin{split}
\lvert \varphi_0 \rangle :=&\ 00\underline{\bf 1}1_2, \qquad \mathrm{idx(\varphi_{0})} = 0 \\
\downarrow& \\
\lvert \varphi_0^{(2)} \rangle :=&\ \underline{\bf 1}001_2, \qquad \mathrm{idx(\varphi_{0}^{(2)})} = C_3^2 = 3, \\
\end{split}
\end{equation*}
and
\begin{equation*}
\begin{split}
\lvert \varphi_0^{(2)} \rangle :=&\ 100\underline{\bf 1}_2, \qquad \mathrm{idx(\varphi_{0}^{(2)})} = C_3^2 = 3 \\
\downarrow& \\
\lvert \varphi \rangle = \lvert \varphi_{0}^{(1)} \rangle :=&\ 1\underline{\bf 1}00_2, \qquad \mathrm{idx(\varphi)} = C_3^2 + C_2^1 = 5.
\end{split}
\end{equation*}
One can refer back to the table to verify the result.
The index in the full Hilbert space, composed from tensor products between basis elements in the two spin sectors, can be computed from the known dimension of the spin-resolved Hilbert spaces.  Given our convention with spin-down configurations occupying the first $N$ LSBs,  $\mathrm{idx}(\varphi) = C_N^{n_{\downarrow}}\cdot\mathrm{idx}_{\uparrow}(\varphi_{\uparrow}) + \mathrm{idx}_{\downarrow}(\varphi_{\downarrow})$, where $\uparrow/\downarrow$ subscripts denote the appropriate spin-restricted subspace. 


%

To be complete, \emph{Paradeisos} requires a ``backward mapping'' for determining a bit sequence from a known index.  This map follows from the accumulated distance between simple neighbors to determine the position of non-zero bits. Pseudocode summarizes the procedure. 
\newline\\
\noindent\begin{tabular}{|l|}
\hline
\textbf{\emph{Paradeisos} backward map}\\
Goal: For $n$ electrons and $N$ sites, map $\mathrm{idx}(\varphi) \mapsto \varphi$\\
\hline
\textbf{function} {\it backward\_map}\,($\mathrm{idx}$; $n$, $N$)\\
\indent \textbf{declare} $\varphi$ = $00 \ldots 00_2$, $m$ = $n$, $p=\mathrm{idx}$\\
\indent \textbf{for} $i = N-1$ to $0$:\\
\indent \indent \textbf{if} $p \geq C_i^m$:\\
\indent \indent\indent set $\varphi$'s $i$-th bit to \texttt{TRUE}\\
\indent \indent\indent $p$\ $\leftarrow$\ $p - C_i^m$\\
\indent \indent\indent $m$\ $\leftarrow$\ $m - 1$\\
\indent \indent \textbf{end}\\
\indent \textbf{end}\\
\indent \textbf{return} $\varphi$\\
\hline
\end{tabular}\\
\newline
\noindent One can verify this procedure in the previous example by inspection.

\section{Mapping within symmetry subspaces}\label{sec:symmetries}

One typically uses symmetries to reduce the dimension of the effective Hilbert space.  These may include $\mathrm{SU(2)}$ spin symmetry~\cite{FehskeBook2008}, point group symmetries such as translational, rotational~\cite{FanoSym1} and inversion symmetry,  or time reversal symmetry that partition the original Hilbert space and block diagonalize the Hamiltonian.   
In the following sections, we illustrate the implementation of \emph{Paradeisos} within both translation and inversion symmetry restricted subspaces.

\subsection{Translational symmetry}

For regular lattice models defined on small clusters with periodic boundary conditions, translation symmetry can lead to a reduction in the Hilbert space dimension by a factor $\sim N$, the total number of lattice sites or unit cells.  Translational invariance typically prompts one to choose a momentum space representation via a discrete fourier transform of the operators $c^{\dagger}_{\bf k}=\frac{1}{\sqrt{N}}\sum_{j} e^{i\textbf{k}\cdot{\bf r}_{j}c^{\dagger}_{j}}$ (similarly for the hermitian conjugate, annihilation operator $c_{\bf k}$).  Transforming the single-band Hubbard model of Eq.~\ref{eq:hubbard} to this momentum space basis leads to 
\begin{equation}
\label{eq:kspacehubbard}
H = \sum_{ \textbf{k}, \sigma}\varepsilon_{\bf k} c^\dagger_{\bf k\sigma} c_{\bf k\sigma}  + \frac{U}{N}\sum_{\bf k,k',q}c^{\dagger}_{\bf k+q\uparrow}
c^{\dagger}_{\bf k'-q\downarrow}c_{\bf k'\downarrow}c_{\bf k\uparrow}
\end{equation}
where $\varepsilon_{\bf k}$ is the kinetic energy in momentum space, typically referred to as the bare band structure.  For the single-band Hubbard model on a square lattice with only nearest, $t$, and next-nearest neighbor, $t'$, hopping terms, $\varepsilon_{\bf k} = -2t(\cos k_x +\cos k_y) -4t^\prime \cos k_x\cos k_y$. 
The Pauli exclusion principle also applies in momentum space; so given a normal ordering of the discrete single-particle momenta $\{{\bf k}_0,\ldots,{\bf k}_{N-1}\}$, bit sequences relate to fermion occupation of single-particle momentum states. Note that the full Hilbert space dimension $D$ in the momentum space representation remains the same as that in the real space representation; however, inspection of the two terms in the Hamiltonian reveals an effective dimensional reduction.  The kinetic term is now diagonal in momentum space, whereas the interaction term scatters fermions of momenta ${\bf k}$ and ${\bf k'}$ to new momenta ${\bf k+q}$ and ${\bf k'-q}$, while leaving the total momentum of the many-body state ${\bf K}$ unchanged.  Thus, ${\bf K}$ can be used to partition the basis elements of the Hilbert space with $D_{\bf K} < D$.

Restriction to a translational subspace is equivalent to fixing the total momentum $\mathbf{K}$, modulo the first BZ. To apply the \emph{Paradeisos} mapping functions in the restricted subspace, one must replace the binomial coefficients by an extended combinatorial $C_{i}^m(\textbf{k})$ following the recursion
\begin{equation}
C_i^m(\textbf{k}) = C_{i-1}^{m-1}(\textbf{k}-\textbf{k}_i) + C_{i-1}^m(\textbf{k}) ,
\end{equation}
where $C_{i}^m(\textbf{k})$ counts the number of ways to arrange $m$ electrons in a state with $i$-bits and a fixed total momentum $\textbf{k}$. The initial conditions for the recursion are
\begin{displaymath}
C_i^0(\textbf{k}) = \left\{ \begin{array}{l}
1 \textrm{, if $\textbf{k}=\textbf{0}\ (\mathbf{\Gamma}-\textrm{point})$}\\
0 \textrm{, otherwise		}
\end{array} \right.
\end{displaymath}
and
\begin{displaymath}
C_i^1(\textbf{k}) = \left\{ \begin{array}{l}
1 \textrm{, if $\textbf{k}=\textbf{k}_i$}\\
0 \textrm{, otherwise}
\end{array} \right.
\end{displaymath}
Following these rules and Eq.~\ref{eq:idx}, an $n$ fermion state with occupation $j_0 < j_1 < \cdots < j_{n-2} < j_{n-1}$ and corresponding total momentum $\mathbf{K} = \textbf{k}_{j_0} + \cdots + \textbf{k}_{j_{n-1}}$ has an index (within the momentum subspace $D_{\bf K}$)
\begin{eqnarray}
\label{eq:basis_index_subspace}
\mathrm{idx}(\varphi) &=& C_{j_0}^1(\textbf{k}_{j_0}) + C_{j_1}^2(\textbf{k}_{j_0}+\textbf{k}_{j_1}) + \dots \nonumber\\
&&\ldots + C_{j_{n-2}}^{n-1}(\mathbf{K}-\mathbf{k}_{j_{n-1}}) + C_{j_{n-1}}^n(\mathbf{K}).
\end{eqnarray}
One can precompute the coefficients $C_{i}^m(\textbf{k})$ using dynamical programming and store them in a lookup-table. A pseudocode description of this forward mapping in the restricted subspace would be 

\noindent\begin{tabular}{|l|}
\hline
\textbf{\emph{Paradeisos} k-space forward map}\\
Goal: For $n$ electrons and $N$ momenta, map \\
$\varphi \mapsto$ $\textrm{idx}(\varphi)$ \\ 
\hline
\textbf{function} {\it forward\_map\_k}\,($\varphi$)\\
\indent \textbf{declare} $\textrm{idx}$ = 0, $m$ = 0, \textbf{k}=0.\\
\indent \textbf{for} $i = 0$ to $N-1$:\\
\indent \indent \textbf{if} $i$-th bit in $\varphi$ is \texttt{TRUE}:\\
\indent \indent \indent $m$\ $\leftarrow$\ $m + 1$\\
\indent \indent \indent $\textbf{k}$\ $\leftarrow$\ mod($\textbf{k} + \textbf{k}_i$,BZ)\\
\indent \indent \indent $\textrm{idx}$\ $\leftarrow$\ $\textrm{idx} + C_{i}^m(\textbf{k})$\\
\indent \indent \textbf{end}\\
\indent \textbf{end}\\
\indent \textbf{return} $\textrm{idx}$, $\textbf{k}$\\
\hline
\end{tabular}
\newline\\
Following a similar procedure as in the full Hilbert space, the momentum restricted subspace backward mapping has a pseudocode

\noindent\begin{tabular}{|l|}
\hline
\textbf{\emph{Paradeisos} k-space backward map}\\
Goal: For $n$ electrons, $N$ momenta, \\ 
and total momentum $\textbf{K}$, map $\textrm{idx}(\varphi) \mapsto \varphi$\\
\hline
\textbf{function} {\it backward\_map\_k}\,($\textrm{idx}$; $n$, $N$, $\textbf{K}$)\\
\indent \textbf{declare} $\varphi$ = $00 \ldots 00_2$, $\textbf{k}$ = $\textbf{K}$, $m$ = $n$, $p=\textrm{idx}$\\
\indent \textbf{for} $i = N-1$ to $0$:\\
\indent \indent \textbf{if} $p \geq C_i^m(\textbf{k})$:\\
\indent \indent\indent set $\varphi$'s $i$-th bit to \texttt{TRUE}\\
\indent \indent\indent $p$\ $\leftarrow$\ $p - C_i^m(\textbf{k})$\\
\indent \indent\indent $m$\ $\leftarrow$\ $m - 1$\\
\indent \indent \indent $\textbf{k}$\ $\leftarrow$\ mod($\textbf{k}-\textbf{k}_i$,BZ)\\
\indent \indent \textbf{end}\\
\indent \textbf{end}\\
\indent \textbf{return} $\varphi$\\
\hline
\end{tabular}\\

The spin-full restricted momentum subspace can no longer be regarded as a direct product space between $\uparrow$ and $\downarrow$ momentum subspaces. Instead, the index of a basis element depends explicitly on both the $\uparrow$ and $\downarrow$ configurations.  Noting the normal ordering convention with the $\downarrow$ bits followed by those for $\uparrow$ right-to-left, the full forward mapping procedure reduces to first performing a forward mapping for the $\downarrow$ portion of the basis element.  The pseudocode for treating the remainder of the forward mapping including the $\uparrow$ portion of a basis element can be written as

\noindent\begin{tabular}{|l|}
\hline
\textbf{\emph{Paradeisos} k-space forward map (spin-full)}\\
Goal: For $n_\uparrow + n_\downarrow$ electrons and $N$ momenta, map \\
$\varphi = \varphi_\uparrow\otimes\varphi_\downarrow \mapsto \textrm{idx}(\varphi)$\\
\hline
\textbf{function} {\it forward\_map\_k\_spin}\,($\varphi_\uparrow$, $\varphi_\downarrow$) \\
\indent \textbf{declare} $\textrm{idx}$ = 0, $m$ = 0, \textbf{k}=0.\\
\indent \textbf{call} {\it forward\_map\_k}\,($\varphi_{\downarrow}$):  $\textrm{idx}_{\downarrow}$, $\textbf{K}_{\downarrow}$; $\textbf{k}$\ $\leftarrow$\ $\textbf{K}_{\downarrow}$.\\
\indent \textbf{for} $i = 0$ to $N-1$:\\
\indent \indent \textbf{if} $i$-th bit in $\varphi_\uparrow$ is \texttt{TRUE}:\\
\indent \indent \indent $m$\ $\leftarrow$\ $m + 1$\\
\indent \indent \indent $\textbf{k}$\ $\leftarrow$\ mod($\textbf{k} + \textbf{k}_i$,BZ)\\
\indent \indent \indent $\textrm{idx}$\ $\leftarrow$\ $\textrm{idx} + \widetilde{C}_{i}^m(\textbf{k}; n_\downarrow)$\\
\indent \indent \textbf{end if}\\
\indent \textbf{end}\\
\indent \textbf{return} $\textrm{idx} + \textrm{idx}_{\downarrow}$ \\
\hline
\end{tabular}\\

\noindent Here one must work with an additional modification of the combinatorial function, which accounts for the spin-down combinations.  This modified version takes the form 
\begin{equation}
\widetilde{C}_i^m(\textbf{k}; n_\downarrow) = \sum_{\bf k_\downarrow}C_i^m(\textbf{k} - \textbf{k}_\downarrow)C_N^{n_\downarrow}(\textbf{k}_\downarrow).
\end{equation}
Again, these quantities may be precalculated and stored in a look-up table.  A similar modification applies for the spin-full backward mapping whose pseudocode becomes
\newline\\
\noindent\begin{tabular}{|l|}
\hline
\textbf{\emph{Paradeisos} k-space backward map (spin-full)}\\
Goal: For $n_\uparrow + n_\downarrow$ electrons, $N$ momenta, \\
and total momentum $\textbf{K}$, map $\textrm{idx}(\varphi) \mapsto \varphi_\uparrow\otimes\varphi_\downarrow$\\
\hline
\textbf{function} {\it backward\_map\_k\_spin}\,($\textrm{idx}$; $n_\uparrow$, $n_\downarrow$, $N$, $\textbf{K}$)\\
\indent \textbf{declare} $\varphi_\uparrow$ = $00 \ldots 00_2$, $\textbf{k}$ = $\textbf{K}$, $m$ = $n_\uparrow$, $p=\textrm{idx}$\\
\indent \textbf{for} $i = N-1$ to $0$:\\
\indent \indent \textbf{if} $p \geq \widetilde{C}_{i}^m(\textbf{k}; n_\downarrow)$:\\
\indent \indent\indent set $\varphi_\uparrow$'s $i$-th bit to \texttt{TRUE}\\
\indent \indent\indent $p$\ $\leftarrow$\ $p - \widetilde{C}_{i}^m(\textbf{k}; n_\downarrow)$\\
\indent \indent\indent $m$\ $\leftarrow$\ $m - 1$\\
\indent \indent \indent $\textbf{k}$\ $\leftarrow$\ mod($\textbf{k}-\textbf{k}_i$,BZ)\\
\indent \indent \textbf{end if}\\
\indent \textbf{end}\\
\indent \textbf{call} {\it backward\_map\_k}\,($p$; $n_\downarrow$, $\textbf{k}$): $\varphi_\downarrow$ \\

\indent \textbf{return} $\varphi = \varphi_\uparrow\otimes\varphi_\downarrow$\\
\hline
\end{tabular}\\


\subsection{Inversion symmetry}

Inversion symmetry ($\mathcal{F}$) implies that a transformation ${\bf r} \to -{\bf r}$ (or ${\bf k} \to -{\bf k}$) leaves the Hamiltonian unchanged. 
To simplify the presentation and without loss of generality, we confine our discussion to one-dimension in real space for a lattice (chain) with $N$ sites indexed sequentially $0, \dots ,N-1$.  Inversion acting on the creation and annihilation operators appearing in the Hamiltonian returns $\mathcal{F}\left(c^{\dagger}_i\right) \to c^{\dagger}_{N-1-i}$ and $\mathcal{F}\left(c_i\right) \to c_{N-1-i}$.  
Since $\mathcal{F}^2 = I$, $\mathcal{F}$ has eigenvalues $\pm 1$, which decomposes the Hilbert space into two orthogonal subspaces, corresponding to symmetric and antisymmetric combinations of the basis elements. 
Note that certain basis elements may already be symmetric in $\mathcal{F}$. 
Based on our conventions, the goal will be to compute the index of a given bit sequence representing the \emph{symmetric} subspace (a similar procedure can be employed for the antisymmetric subspace).

For $n$ electrons on $N$ sites, partition the representatives into three classes according to their bits in positions $N-1$ and $0$:
\begin{itemize}
\item[A:] \quad $ \lvert \varphi_{A} \rangle := \overset{N-1}{0} \dots \overset{i=0}{1_2}$ 
\item[B:] \quad $ \lvert \varphi_{B} \rangle := \overset{N-1}{0} \dots \overset{i=0}{0_2}$ 
\item[C:] \quad $ \lvert \varphi_{C} \rangle := \overset{N-1}{1} \dots \overset{i=0}{1_2}$ 
\end{itemize}
The symmetric states in class A are of the form
\begin{equation*}
\lvert \varphi_{A}^{\mathrm{s}} \rangle := \big( 0\ a_{N-2} \dots a_1\ 1_2 + 1\ a_1\ \dots a_{N-2}\ 0_2\big) /\sqrt{2},
\end{equation*}
which obviates the need for bit sequences of the form $\overset{N-1}{1} \dots \overset{i=0}{0_2}$ that are included by construction.  Within class A, the index can be determined by standard \emph{Paradeisos} (forward mapping without symmetry). The index for elements in classes B and C can be determined by recursive partitioning for bits in positions $N-d$ and $d-1$, where $d$ is the recursion depth.  For each recursion step the same classification scheme can be employed, which results in a tree with class A, B, or C nodes. Figure~\ref{fig:inversion_classification_tree} shows such a tree to illustrate this process.
%

\tikzstyle{level 1}=[level distance=2.0cm, sibling distance=4.5cm]
\tikzstyle{level 2}=[level distance=2.0cm, sibling distance=1.5cm]
\tikzstyle{level 3}=[level distance=2.25cm, sibling distance=0.65cm]

\tikzstyle{bag} = [text width=1em, text centered]
\tikzstyle{end} = [circle, minimum width=1pt, fill, inner sep=0pt]

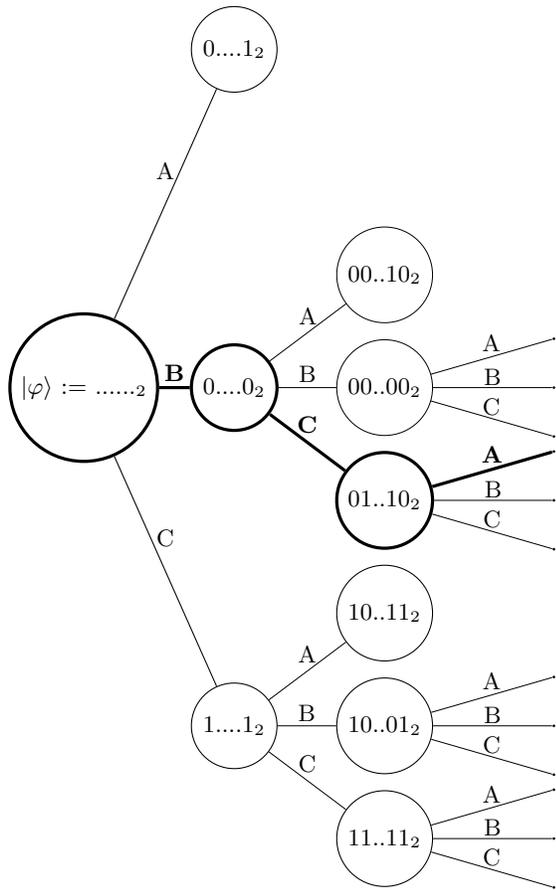
\begin{figure}[ht]
\begin{tikzpicture}[grow=right]
\node[draw, circle, very thick] {$\lvert \varphi \rangle$ := ......$_2$}
    child {
        node[draw, circle] {1....1$_2$}       
                 child {
                    node[draw, circle] {11..11$_2$}    
                    	 	child {
                                node[end, label=right:] {}
                                edge from parent [very thin]
                                node[above, inner sep=0.15em] {C}    
                             }
                             child {
                                node[end, label=right:] {}
                                edge from parent [very thin]
                                node[above, inner sep=0.1em] {B}    
                             }
                             child {
                                node[end, label=right:] {}
                                edge from parent [very thin]
                                node[above, inner sep=0.2em] {A}    
                             }
                    edge from parent [very thin]
                    node[above] {C}
                }
                child {
                    node[draw, circle] {10..01$_2$}  
                    	 	child {
                                node[end, label=right:] {}
                                edge from parent [very thin]
                                node[above, inner sep=0.15em] {C}    
                             }
                             child {
                                node[end, label=right:] {}
                                edge from parent [very thin]
                                node[above, inner sep=0.1em] {B}    
                             }
                             child {
                                node[end, label=right:] {}
                                edge from parent [very thin]
                                node[above, inner sep=0.2em] {A}    
                             }
                    edge from parent [very thin]
                    node[above, inner sep=0.2em] {B}    
                }
                child {
                    node[draw, circle] {10..11$_2$} 
                    edge from parent [very thin]
                    node[above] {A}
               }
        edge from parent 
        	    node[above, inner sep=1em] {C}
    }
    child {
        node[draw, circle, very thick] {0....0$_2$}        
        child {    
            node[draw, circle, very thick] {01..10$_2$}        
                 child {
                    node[end, label=right:
                        {}] {}
                    edge from parent [very thin]
                    node[above, inner sep=0.15em] {C}
                }
                child {
                    node[end, label=right:
                        {}] {}
                    edge from parent [very thin]
                    node[above, inner sep=0.1em] {B}
                }
                child {
                    node[end, label=right:
                        {}] {}
                    edge from parent [very thick]
                    node[above, inner sep=0.2em] {$\mathbf{A}$}
                }
            edge from parent  [very thick]       
                node[above] {\bf{C}}  
        }
            child {
                node[draw, circle, very thin] {00..00$_2$} 
                    	 	child {
                                node[end, label=right:] {}
                                edge from parent [very thin]
                                node[above, inner sep=0.15em] {C}    
                             }
                             child {
                                node[end, label=right:] {}
                                edge from parent [very thin]
                                node[above, inner sep=0.1em] {B}    
                             }
                             child {
                                node[end, label=right:] {}
                                edge from parent [very thin]
                                node[above, inner sep=0.2em] {A}    
                             }
                edge from parent [very thin]
                node[above, inner sep=0.2em] {B}
            }
            child {
                node[draw, circle, very thin] {00..10$_2$} 
                edge from parent [very thin]
                node[above] {A}
            }
        edge from parent  [very thick]        
            node[above, inner sep=0.2em] {\bf{B}}
    }
    child {
        node[draw, circle] {0....1$_2$}        
        edge from parent         
            node[above, inner sep=1em] {A}
    };
\end{tikzpicture}
\caption{A tree up to depth $d=4$ (bits not shown for brevity) in the recursive classification of a bit sequence representing a basis element in the symmetric subspace.  Branches typically terminate at an A class bit sequence from which the standard \emph{Paradeisos} forward mapping can be used to obtain the index of the state at that recursion depth. The highlight shows the path to a state such as $010110_2$, for example.}
\label{fig:inversion_classification_tree} 
\end{figure}

There are a few things to note about this construction.  First, each class A node is a leaf node (which contains a set of states) -- the recursion ends and \emph{Paradeisos} forward mapping can be used to determine the index at that depth.  Leaf nodes can exist for class B and C, but only if the maximum recursion depth has been reached based on the number of bits $N$ in each sequence.  Those leaves that end with B or C have a palindrome pattern, containing only one state. The height of the tree is $N/2 + 1$ for $N$ even or $(N+1)/2$ for $N$ odd, given that the number of bits yet to be classified at each recursion step decreases by $2$. 

By convention, the indices for class A are the smallest, followed by those for class B, and lastly those for class C, at the same depth on the classification tree. 
In this way, the standard order has been destroyed, and replaced by a set of continuous indices for all the representative elements in the symmetric subspace. One observes that class A contains $C_{i-2}^{m-1}$ representatives (the same number for either the symmetric or antisymmetric subspaces by construction) for $i$ ``active'' bits and $m$ non-zero bits at the current depth, as the inner bit sequence has no bearing on the A, B, or C classifications.  Similarly, the number of representatives at a particular level in class B or C can be determined recursively by $S_{i}^{m}$ =  $C_{i-2}^{m-1}$ + $S_{i-2}^{m}$ + $S_{i-2}^{m-2}$, where $S_{i}^{m}$ is the total number of representative elements in the symmetric subspace for $i$ total bits and $m$ non-zero bits.  The three terms represent the number of elements for class A, B and C, respectively, at the next lower level. $S_{i}^{m}$ can be calculated in a bottom-up manner using dynamical programming. 
In this way, the number of preceding basis elements can be calculated by a depth-first search of the tree, counting $C_{i-2}^{m-1}$ elements for a class A leaf and a single basis element for class B or C leaves. 

\noindent\begin{tabular}{|l|}
\hline
\textbf{\emph{Decision Tree} for inversion symmetry}  \\
Goal: construct root and node \emph{structure} \\
\hline
@ depth $d$: \\
\indent node.$\varphi_L$ ($d$ leftmost bits in $\varphi$) \\
\indent node.$\varphi_R$ ($d$ rightmost bits in $\varphi$) \\
\indent node.count (number of representatives) \\
\indent node.A (pointer to depth $d+1$, class A) \\
\indent node.B (pointer to depth $d+1$, class B) \\
\indent node.C (pointer to depth $d+1$, class C) \\
\indent node.parent (pointer to depth $d-1$) \\
\hline
\textbf{function} {\it decision\_tree}\,($n$, $N$)\\
\indent \textbf{for} $d = N/2$ to $0$: ($N$ even) \\
\indent \indent \textbf{construct} \emph{root} and \emph{nodes} @ $d$ \\ 
\indent \indent \indent \emph{assign} node.$\varphi_L(\varphi_R)$ \\ 
\indent \indent \indent \emph{assign} node.A(B,C) \\
\indent \indent \indent \emph{assign} node.parent \\
\indent \indent \textbf{calculate} node.count \\
\indent \indent \indent $C_i^m$ (class A) \\
\indent \indent \indent $S_{i}^{m}$ (class B or C) \\
\indent \textbf{end} \\
\indent \textbf{return} {\it root} \\
\hline
\end{tabular}
\newline

Unfortunately, one cannot simultaneously obtain the benefits of using both translation and inversion symmetry in \emph{Paradeisos}. 
Invoking translation symmetry typically reduces the effective Hilbert space dimension much more than the reduction obtained from the inversion symmetric subspace.  However, the inversion symmetric algorithm may be employed for some problems that lack translation invariance, such those with open boundaries.

To be complete, we present pseudocode for the forward and backward mapping \emph{Paradeisos} algorithms with inversion symmetry.

\noindent\begin{tabular}{|l|}
\hline
\textbf{\emph{Paradeisos} inversion forward map}\\
Goal: For $n$ electrons and $N$ sites, 
map $\varphi \mapsto \mathrm{idx}(\varphi) $\\
\hline
\textbf{function} {\it forward\_map\_inversion}\,($\varphi$)\\
\indent node $\leftarrow$ {\it decision\_tree}\,($n$, $N$) \\
\indent \textbf{for} $d = 0$ to $N/2 - 1$: ($N$ even)\\
\indent \indent \textbf{switch}($(N-1-d)$-th bit in $\varphi$, $d$-th bit in $\varphi$): \\
\indent \indent \textbf{case: }  \texttt{FALSE}, \texttt{TRUE}:\\
\indent \indent \indent $\phi$ $\leftarrow$ ($d$+1)-th through $(N-2-d)$-th bits in $\varphi$ \\
\indent \indent \indent $\textrm{idx}$\ $\leftarrow$\ $\textrm{idx}$ + {\it forward\_map}($\phi$) \\
\indent \indent \indent \textbf{return} $\textrm{idx}$ \\
\indent \indent \textbf{case: } \texttt{FALSE}, \texttt{FALSE}:\\
\indent \indent \indent $\textrm{idx}$\ $\leftarrow$\ $\textrm{idx}$ + node.A.count \\
\indent \indent \indent node\ $\leftarrow$\ node.B \\
\indent \indent \textbf{case:} \texttt{TRUE}, \texttt{TRUE}:\\
\indent \indent \indent $\textrm{idx}$\ $\leftarrow$\ $\textrm{idx}$ + node.A.count + node.B.count \\
\indent \indent \indent node $\leftarrow$ node.C \\
\indent \textbf{end}\\
\indent \textbf{return} $\textrm{idx}$ \\
\hline
\end{tabular}\\
\newline


\begin{figure*}[ht]
\centering
\includegraphics[width=1.8\columnwidth]{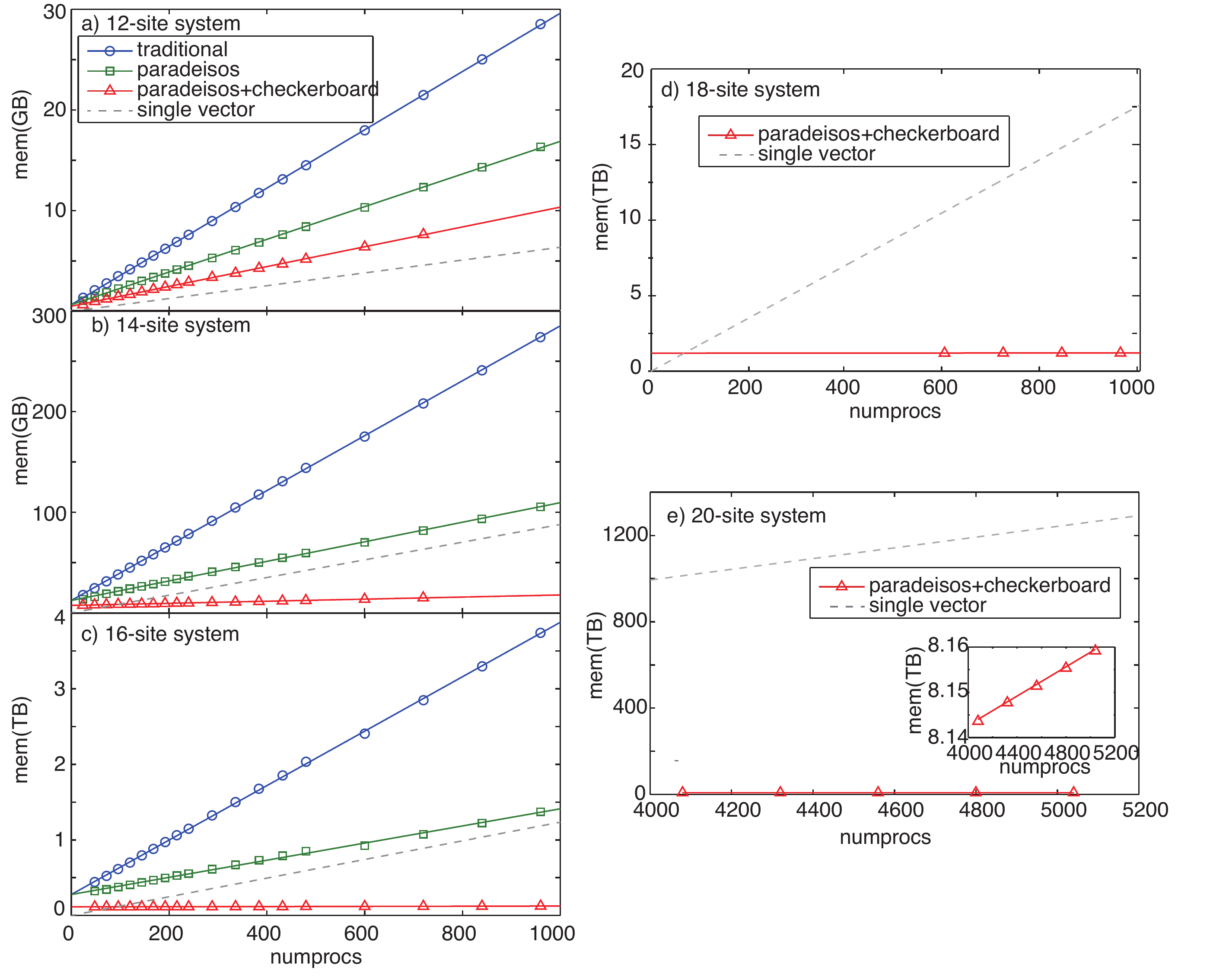}
\caption{(Color online) Comparison of aggregate memory usage as a function of the number of processors for sparse matrix construction and diagonalization (obtaining the groundstate wavefunction), for a one-dimensional spin-1/2 Hubbard model in real space without invoking any Hilbert space reductions due to symmetry.  The blue curves correspond to traditional binary search, green curves correspond to \emph{Paradeisos}, and red curves relate to \emph{Paradeisos} combined with a checkboard decomposition. The dashed lines indicate the memory required to store a single vector within the Hilbert space serially for each process. Note significant differences in scale with Hilbert space dimension (lattice size).}
\label{fig:RealspaceMem}
\end{figure*}

\noindent
\begin{tabular}{|l|}
\hline
\textbf{\emph{Paradeisos} inversion backward map}\\
Goal: For $n$ electrons and $N$ sites, 
map  $\mathrm{idx}(\varphi) \mapsto \varphi$\\
\hline
\textbf{function} {\it backward\_map\_inversion}\,(idx; $n$, $N$)\\
\indent node $\leftarrow$ {\it decision\_tree}\,($n$, $N$), $p = $ idx \\
\indent \textbf{for} $d = 0$ to $N/2 - 1$: ($N$ even)\\
\indent \indent \textbf{if} $p$ $\geq$ node.A.count + node.B.count:  \\
\indent \indent\indent $p$ $\leftarrow$ $p$ $-$ node.A.count $-$ node.B.count\\
\indent \indent\indent node $\leftarrow$ node.C\\
\indent \indent \textbf{elseif} $p$ $\geq$ node.A.count:  \\
\indent \indent\indent $p$ $\leftarrow$ $p$ $-$ node.A.count\\
\indent \indent\indent node $\leftarrow$ node.B\\
\indent \indent \textbf{else}:  \\
\indent \indent \indent node $\leftarrow$ node.A \\
\indent \indent \indent $m$ $\leftarrow$ $n$ $-$ bit\_count(node.$\varphi_L$ $\wedge$ node.$\varphi_R$, 1) \\
\indent \indent \indent $\phi$ $\leftarrow$ {\it backward\_map}($p$; $m$, $N - 2d$) \\
\indent \indent\indent \textbf{return} node.$\varphi_L$ $\wedge$ $\phi$ $\wedge$ node.$\varphi_R$ \\
\indent \indent \textbf{end if}\\
\indent \textbf{end}\\
\indent \textbf{return} node.$\varphi_L$ $\wedge$ node.$\varphi_R$ \\
\indent ($\wedge$ represents concatenation of binary sequences) \\
\hline
\end{tabular}

\section{Numerical Benchmarks}

The goal in devising the \emph{Paradeisos} algorithm was to address one of the remaining bottlenecks for sparse matrix eigensolvers -- memory requirements for basis element hashing in traditional binary search that lead to a large communications overhead or significant aggregate memory consumption in large-scale parallel implementations.  In this section we benchmark the performance of \emph{Paradeisos} against binary search, and investigate additional benefits when coupling the algorithm with the checkerboard decomposition for data parallelism.  We concentrate specifically on how memory usage scales with the number of processors for single-band Hubbard model calculations on clusters of varying size, as the scaling slope usually determines the quality of a parallelization scheme.

\begin{figure}[ht]
\centering
\includegraphics[width=0.9\columnwidth]{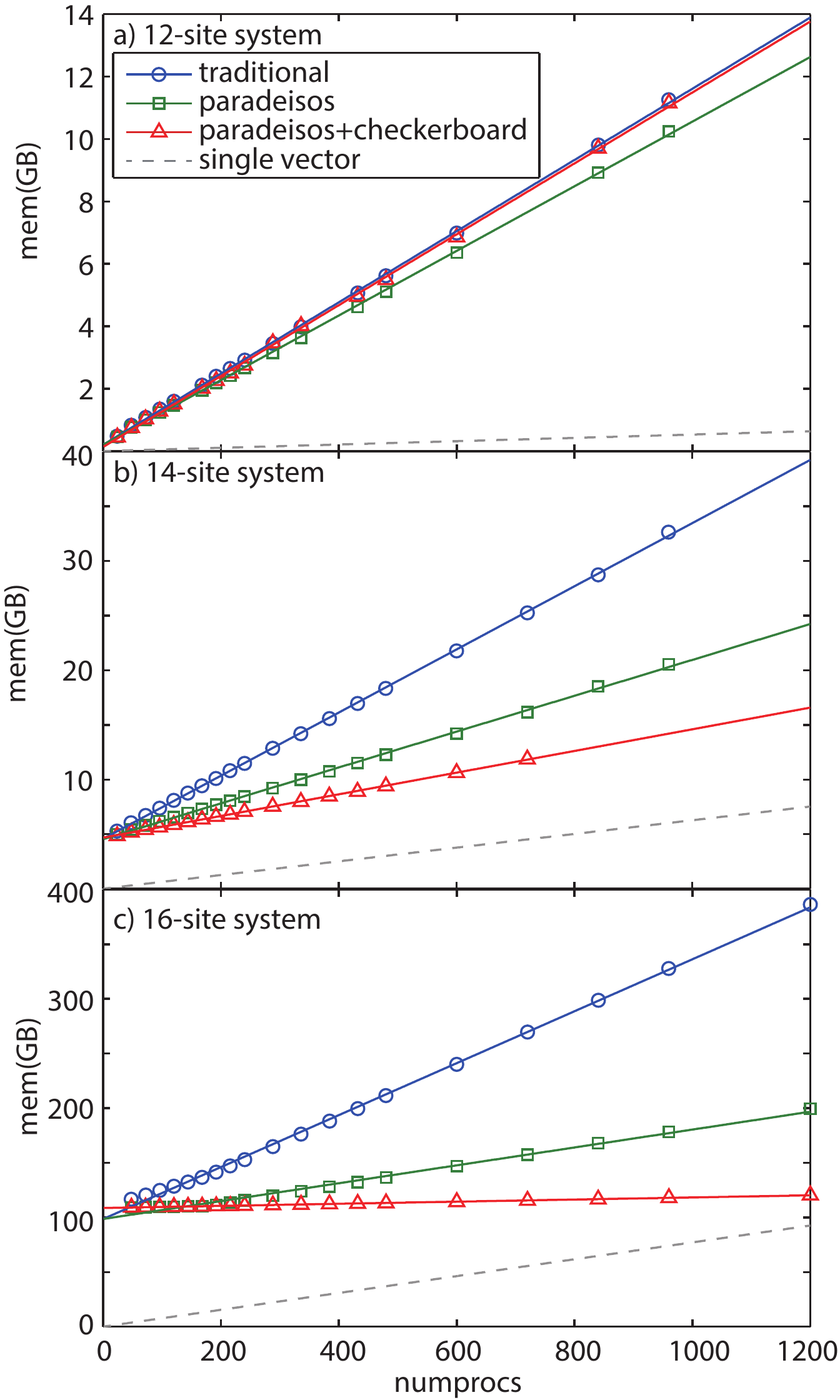}
\caption{(Color online) Comparison between traditional binary search (blue), \emph{Paradeisos} (green), and \emph{Paradeisos} combined with a checkerboard decomposition (red).  Each curve shows overall memory consumption for a one-dimensional spin-1/2 Hubbard model on chains (lattices) of length (a) 12, (b) 14, and (c) 16 sites.  Each one represents the memory required for solving the problem in the $K = 0$ momentum subspace, which contains the groundstate wavefunction.}
\label{fig:KspaceMem}
\end{figure}

In a parallel matrix-vector algorithm, the storage of a single vector of size $D$ on each processor characterizes the typical memory cost and can usually to be used as a reference to determine the necessity of a sophisticated parallelization scheme like the ``checkerboard" decomposition. Thus, we denote such a memory cost as grey lines in Figs.~\ref{fig:RealspaceMem} and ~\ref{fig:KspaceMem}.  As shown in Fig.~\ref{fig:RealspaceMem}, the traditional binary search method (blue) has a relatively large scaling slope with system size, making the parallelization cost extremely expensive for large cluster calculations. Once the overhead cost exceeds the benefits of parallelization, adding more processors would not help to solve the memory problem.  \emph{Paradeisos} (green) succeeds in reducing the slope. 
However, the steady increase of aggregate memory with the number of processors still limits parallelization.  When one applies \emph{Paradeisos} with the checkerboard decomposition (red), the situation changes dramatically. The scaling slope becomes relatively small and shows little change with growth in the dimension of the Hilbert space (dictated by the lattice size).  The memory overhead remains effectively constant, an important consideration for parallel computations that then may be cheaply divided over many nodes or processors. For the largest problems, \emph{Paradeisos} improves the scaling beyond that of even storing a single vector in the Hilbert space for each process.  

\begin{figure}[ht]
\centering
\includegraphics[width=0.8\columnwidth]{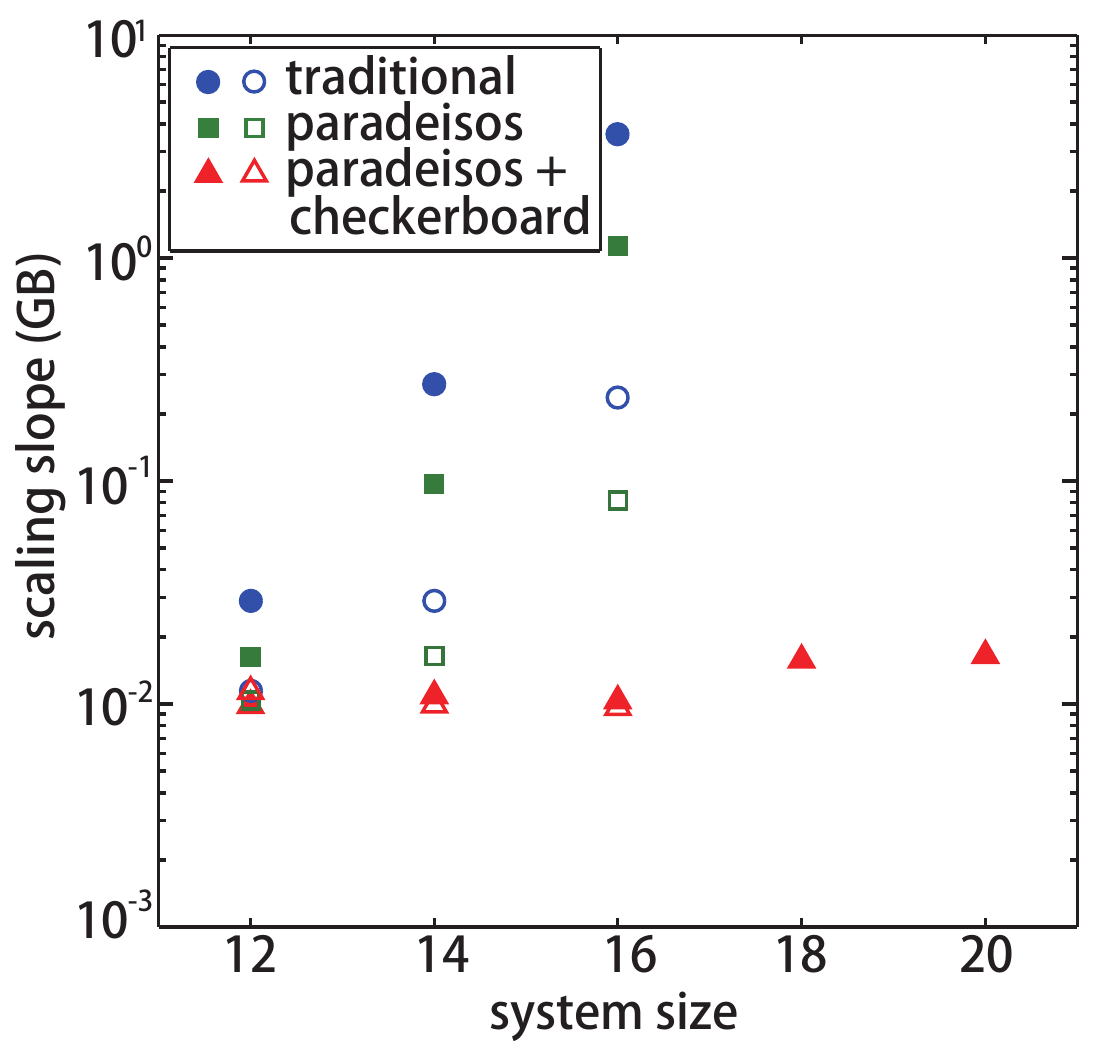}
\caption{Summary of the memory scaling slopes as a function of system size from Figs.~\ref{fig:RealspaceMem} and \ref{fig:KspaceMem}. The solid dots represent results from the real space implementation, while the open symbols correspond to those obtained from invoking the translation symmetry reduction of the effective Hilbert space.}
\label{fig:slope}
\end{figure}

As shown in Fig.~\ref{fig:KspaceMem}, when applying symmetry reduction for small Hilbert space dimensions (the smallest lattice problems), \emph{Paradeisos} alone performs marginally better than binary search and even when combined with the checkerboard decomposition, primarily due to residual memory overhead from other parts of the sparse matrix diagonalization code.  With increasing Hilbert space dimension (lattice size), the results remain qualitatively similar to the those from the real space implementation of the algorithm without symmetry reduction.  Fig.~\ref{fig:slope} compares the memory scaling slope for various lattice sizes, showing the significant improvements in memory consumption that can be realized by combining \emph{Paradeisos} with the checkerboard decomposition. 


\section{Summary and Discussion}
We have proposed a perfect hashing algorithm -- a direct mapping between Hilbert space basis elements and their corresponding index -- for use in sparse matrix eigenvalue problems. Compared to the traditional binary search, the present algorithm provides a considerable savings in aggregate memory usage without incurring additional penalties to the time complexity.  Moreover, in concert with a checkerboard decomposition scheme, the memory overhead can be negligible (effectively independent of problem size), implying efficient parallelization for large size compute environment.  The Paradeisos algorithm is also compatible with additional point group symmetries -- which can highly reduce the Hilbert space dimension, and thus can be efficiently applied to most quantum many-body systems or models. The algorithm can be extended to apply on many-body bosonic basis states, at which a particle number truncation is necessary. The algorithm eschews storage, and as such, may be utilized in \emph{matrix-free} implementations of eigenvalue solvers.

\section{Acknowledgements}
\begin{acknowledgments}
This work was supported at SLAC and Stanford University by the U.S.~Department of Energy, Office of Basic Energy Sciences, Division of Materials Sciences and Engineering, under Contract No.~DE-AC02-76SF00515. Y.W.\ was supported by the Stanford Graduate Fellows in Science and Engineering, and C.M.\ acknowledges support from the Alexander von Humboldt foundation via a Feodor Lynen fellowship. The computational work was performed using the resources of the National Energy Research Scientific Computing Center supported by the U.S. Department of Energy, Office of Science, under Contract No.~DE-AC02-05CH11231.
\end{acknowledgments}

\bibliography{references}

\end{document}